\documentclass[aps,prc,twocolumn,showpacs]{revtex4}
\usepackage{amssymb}
\usepackage{color}
\usepackage{ulem}
\usepackage{color}
\usepackage{graphicx}
\usepackage{epsfig}
\usepackage{bm}
\def\be {\begin{equation}}
\def\ee {\end{equation}}
\def\nn {\nonumber}
\def\bea {\begin{eqnarray}}
\def\eea {\end{eqnarray}}

\def\e{\epsilon}

\newcommand{\ep}{\epsilon}
\newcommand{\om}{\omega}  
\newcommand{\vk}{\vec k}

\newcommand{\Ds}{D \!\!\!\! /}

\def\zbf#1{{\bf {#1}}}

\def\vec#1{\mathchoice
        {\mbox{\boldmath $#1$}}
        {\mbox{\boldmath $#1$}}
        {\mbox{\boldmath $\scriptstyle #1$}}
        {\mbox{\boldmath $\scriptscriptstyle #1$}}
}

\begin{document}




\title{Impact of magnetic field on shear viscosity of quark matter in Nambu-Jona-Lasinio model}
%
\author{Sabyasachi Ghosh$^1$, Payal Mohanty$^2$, Bhaswar Chatterjee$^3$, Arghya Mukharjee$^4$, Hiranmaya Mishra$^5$}

\affiliation{$^1$ Indian Institute of Technology Bhilai, GEC Campus, Sejbahar, Raipur-492015,
Chhattisgarh, India}

\affiliation{$^2$ National Institute of Science Education and Research, HBNI, 752050 Odisha, India.}
\affiliation{$^3$ Department of Physics, Indian Institute of Technology Roorkee, 
Roorkee 247 667, India}
\affiliation{$^4$ Saha Institute of Nuclear Physics, 1/AF Bidhannagar, 
Kolkata 700064, India}
\affiliation{$^{5}$ Theory Division, Physical Research Laboratory, 
Navrangpura, Ahmedabad 380 009, India}

\begin{abstract}
We have investigated shear viscosity of 
quark matter in presence of a strong uniform magnetic
field background where Nambu-Jona-Lasinio model has been considered to describe the 
magneto-thermodynamical properties of the medium. 
In presence of magnetic field, shear viscosity coefficient 
gets split into different components because of anisotropy
in tangential stress of the fluid. Four different components
can be merged to two components in limit of strong field, where
collisional width of quark becomes much lower than its synchrotron
frequency. A simplified contact diagram of quark-quark interaction
can estimate a small collisional width, where strong field limit
expressions are exactly applicable. Although, for RHIC or LHC matter,
one can expect a large thermal width, for which generalized four components
viscosities are necessary. We have explored these all different 
possible cases in the thermodynamical framework of Nambu-Jona-Lasinio model. 


%
\end{abstract}

%
%
%
%




\maketitle
\section{Introduction}
One of the major update in the research of heavy ion collision (HIC) experiments
like RHIC and LHC is that the produced medium behaves like
a nearly perfect fluid~\cite{Schafer_Rev}, with smallest shear viscosity to 
entropy density ratio ($\eta/s$), ever observed in nature.
On the other hand, recent progress in the HIC research have speculated that 
the produced medium may also be subjected to
a strong magnetic field~\cite{Tuchin_Rev} in the non-central heavy-ion collisions. The possible space-time dependence
of this produced magnetic field has been investigated in 
Refs.~\cite{{Skokov:2009qp},{Voronyuk:2011jd},{Bzdak:2011yy},{Deng:2012pc},{Deng:2014uja}}. 
A considerable  amount of research work has already been performed in  understanding the influence of the magnetic field 
on the QCD phase diagram.  See, for example, the review article~\cite{jens_rmp} for recent updates.
The modification of the QCD phase diagram in presence of magnetic field is directly related to the corresponding 
change in the quark condensate and its enhancement with magnetic
field is known as magnetic catalysis (MC) which is quite expected feature
in vacuum as well as at finite 
temperature~\cite{{Miransky},{Marco1},{Marco2},{Boomsma:2009yk},{BC_HM1},{BC_HM2}}. 
However, recent calculations, based on lattice quantum chromodynamics (LQCD)~\cite{Bali,Bornyakov}
have found inverse magnetic catalysis, whose possibility is also indicated
by some effective QCD model calculations~\cite{Ayala_LSM1,Ayala_LSM2,Krein1,Krein2}.
The modifications pertaining to the QCD phase diagram may also have some impact in the 
transport properties of the medium produced in HIC.   
In presence of magnetic field, different transport coefficients like shear
viscosity~\cite{Li_shear,Nam_shear,Sedarkian_shear,Tawfik_shear,Tuchin_shear,
Hattori_bulk,Sedarkian_bulk,Huang_bulk},
bulk viscosity~\cite{Hattori_bulk,Sedarkian_bulk,Huang_bulk,Agasian_bulk1,Agasian_bulk2}
and electrical conductivity~\cite{Nam_cond,Hattori_cond1,Hattori_cond2,Sedarkian_cond,
Kerbikov_cond,Feng_cond,Buividovich_cond,Fukushima_cond} of quark matter are 
calculated in recent times. The simulation of magnetohydrodynamics~\cite{Roy:2015kma,Pu:2016ayh} 
as well as the transport simulation for an external magnetic field~\cite{Das:2016cwd}
may require these temperature and magnetic field dependent transport coefficients for their 
future up-gradation.

Among the different transport coefficients, only the  shear viscosity is  our matter of
interest in the  present work, where
two flavor Nambu-Jona-Lasinio (NJL) model has been used as a dynamical framework.
Among the earlier calculations of shear viscosity for magnetized 
matter~\cite{Li_shear,Nam_shear,Sedarkian_shear,Tawfik_shear,Tuchin_shear,
Hattori_bulk,Sedarkian_bulk,Huang_bulk,Ashutosh}, we find that 
Refs.~\cite{Li_shear,Nam_shear,Sedarkian_shear,Tawfik_shear} have not
explored its component decomposition, which is explicitly analyzed in 
Refs.~\cite{Tuchin_shear,Hattori_bulk,Sedarkian_bulk,Huang_bulk,Ashutosh}.
This component decomposition of shear viscosity due to anisotropy, 
created by external magnetic field or other sources, is well studied
in the direction of gauge gravity duality 
(See~\cite{Finazzo_shear,Jain} and references therein). 

We have first followed the strong field limit expression, obtained 
in Refs.~\cite{Tuchin_shear,Landau}, where four components of shear viscosity
merge to two main components, as also found in gauge gravity dual theory~\cite{Finazzo_shear,Jain}.
One is normal-type shear viscosity and another is Hall-type component. Normal component
depends on both collisional and synchrotron frequencies, but Hall component depends completely on synchrotron
frequency in the strong field limit. However, below that strong field limit, both the components
can depend on both frequencies. We have also studied on general structure of four different
components in the moderate field zone, which is expected in RHIC or LHC experiments. 


The article is organized as follows. In Sec.~(\ref{sec2}), the background 
formalism of NJL model is addressed. Next Sec.(\ref{sec3}) cover the formalism of shear viscosity for 
the case of strong  magnetic field in subsection.(\ref{sec3A}) and then its
corresponding numerical outcome in subsection.(\ref{sec3B}). Realizing strong field limit
can not be applicable for RHIC or LHC matter, which should have small collisional relaxation time, we have
gone through strong field case to general case, whose modified formalism and corresponding 
numerical outcome is discussed in Sec.~(\ref{sec4A}) and (\ref{sec4B}) respectively.
At the end, investigations of all these different possible cases are summarized in Sec.~(\ref{sec6}).   

\section{NJL model in presence of magnetic field}
\label{sec2}
We shall consider here, two flavor (u, d quarks)  NJL model with a determinant interaction with 
the Lagrangian density given as~\cite{Frank,Boomsma:2009yk}
\bea
{\cal L} &=&
\bar\psi(i\Ds-m)\psi 
\nn\\
&&+ G\sum_{a=0}^3\left[(\bar\psi\tau^a\psi)^2 + (\bar\psi i\gamma_5\tau^a\psi)^2\right]
\nn\\
&&+ K\left[det_f\bar\psi(1+\gamma_5)\psi + det_f\bar\psi(1-\gamma_5)\psi\right]~,
\label{lag2fl}
\eea
where $\psi=(u,~ d)^T$ is the doublet of quarks, $m=(m_u,~ m_d)$ is the current quark mass
with $m_u=m_d$.
The first term is basically the Dirac Lagrangian  
in presence of an external magnetic field, which we assume to be constant and in
the direction of $z$-axis.
For calculational purpose, we shall further choose the gauge such that the corresponding 
electromagnetic potential is given by
 $A_\mu(\vec x)=(0,~ 0,~ Bx,~ 0)$.  The second line is attractive part of the quark 
 anti-quark channel of the Fiertz transformed color
current-current interaction. The third line is the 't-Hooft determinant 
interaction in the flavor space that describes the
 effects of instantons and is flavor mixing. $\tau^a, a=0\cdots 3$ are the U(2) generators in the flavor space. 
In the absence of magnetic field the interaction is invariant
under 
$SU(2)_L\times SU(2)_R\times U_V(1)$. 
The second term  has an additional $U(1)_A$ 
symmetry while the t-Hooft term
does not have this symmetry and reflects the $U(1)_A$ anomaly of QCD.

The thermodynamic potential corresponding to Eq.(\ref{lag2fl}) can be computed exactly 
in the same manner as was done previously in 
Ref.~\cite{BC_HM1}, that was done for three flavors in a variational 
method with an explicit structure for the vacuum with 
quark anti-quark condensates. The thermodynamic potential is then given as
\bea
\Omega&=&\sum_i\Omega_0^i+\sum_i\Omega^i_{field}+\sum_i\Omega^i_{med}
\nn\\
&&~~~~~~~~~~~~+2G\sum_i {I_s^{i}} ^2+2KI_s^uI_s^d
\label{thpot}
\eea
where, i is the flavor index. 
We might mention here that the above thermodynamic potential can also be
derived in a mean field approximation \cite{Frank}.
The vacuum term for i-th flavor $\Omega_0^i$ is given as
\bea
\Omega_0^i &=& -\frac{2N_c}{(2\pi)^3}\int d\zbf p\sqrt{\zbf p^2+ M_i^2}\theta(\Lambda-|\zbf p|)\nonumber\\
&=&-\frac{N_c}{8\pi^2}\bigg[\Lambda \sqrt{\Lambda^2+M_i^2}(2\Lambda^2+M_i^2)
\nonumber\\
&& -M_i^4\log\frac{\Lambda+\sqrt{\Lambda^2+M_i^2}}{M_i}\bigg],
\label{vacomg}
\eea
with, $\Lambda$ as the three momentum cutoff associated with the NJL model.
The field contribution that arises from the effect of magnetic field on the Dirac vacuum is given by
\bea
\Omega_{field}^i &=& -\frac{N_c}{2\pi^2}\sum_i|q_iB|^2\bigg[\zeta^\prime(-1,x_i) 
\nonumber\\
&& -\frac{1}{2}(x_i^2 - x_i)\ln{x_i} + \frac{x_i^2}{4}\bigg], 
\eea
where we have defined a dimension less quantity,
$x_i = M_i^2/2|q_iB|$ , i.e. the mass parameter in units of magnetic field 
and $\zeta^\prime(-1,x) = d\zeta(z,x)/dz|_{z = 1}$ 
is the derivative of the Riemann-Hurwitz $\zeta$ function which is given by 
\bea
&&\zeta^\prime(-1,x) = \frac{\ln x}{2}\left[x^2 - x + \frac{1}{6}\right] - 
\frac{x^2}{4}
\nonumber\\
&&~+ x^2 \int_0^\infty\frac{2\tan^{-1}y + y\ln(1 + y^2)}{e^{2\pi xy} - 1}dy.
\label{derivz}
\eea
Finally, the medium contribution $\Omega_{med}^i$ is given as
\be
\Omega_{med}^i = \frac{-N_c}{\pi^2}\sum_{n=0}^{n_{max}}\frac{\alpha_n|q_i B|}{\beta}
\int dp_z\log (1+e^{-\beta\omega^i_{n}})
\ee
with the single particle energy in presence of magnetic field  $\omega^i_n=\sqrt{p_z^2+2n|q_i|B+m^2}$. 
The condition of a sharp
three momentum cutoff translates to a finite number of Landau level summation with 
$n_{max}=Int[\frac{\Lambda^2}{2|q_i|B}]$ when $p_z=0$. Further,
for the medium contributions this also leads to a cutoff for the $|p_z|$ 
as $\Lambda^\prime=\sqrt{\Lambda^2-2n|q_i|B}$
for a given value of $n$.

Similarly, in Eq.(\ref{thpot}), the quark condensate $I_s^i=-\langle\bar\psi^i\psi^i\rangle$, 
can be separated into a zero field vacuum term, a finite field dependent term 
and a medium dependent term as
\bea
I_s^i &\equiv& -\langle\bar\psi^i\psi^i\rangle = \frac{2N_c}{(2\pi)^3}\int_{|\zbf p|<\Lambda} 
{d\vec p}\frac{M^i}{\sqrt{\vec p^2 + {M^i}^2}}
\nonumber\\
&+& \frac{N_c M^i|q^iB|}{(2\pi)^2}\left[x^i(1-\ln{x^i}) 
\right.\nn\\
&&\left. ~~~~~~~~~~~~~~~~+ \ln{\Gamma(x^i)} +
\frac{1}{2}\ln{\frac{x^i}{2\pi}}\right]
\nonumber\\
&-&\sum_{n=0}^{n_{max}}\frac{N_c|q^i|B\alpha_n}{(2\pi)^2}
\int dp_z\frac{M^i}{\omega_i^n}\frac{1}{1+\e^{-\beta\omega_i^n}}
\nonumber\\
&=& {I_s^i}_{vac} + {I_s^i}_{field}+{I_s^i}_{med}.
\label{Isif}
\eea
\noindent
The zero field vacuum contribution, ${I_s^i}_{vac}$, can be analytically calculated using a sharp 
momentum cutoff $\Lambda$ and can be written as
\bea
I_{s_{vac}}^{i}&=&\frac{N_c M_s^i}{2\pi^2}\left[\Lambda\sqrt{\Lambda^2+{M^i}^2}
\right.\nn\\
&&\left.~~~~~-{M^i}^2\log\left\lbrace
\frac{\Lambda+\sqrt{\Lambda^2+{M^i}^2}}{M^i}\right\rbrace\right]~.
\label{Isvac}
\eea

The constituent quark mass $M^i$ satisfies the gap equation
\be
M_i=m_i+4GI_s^i+2K|\epsilon^{ij}|I_s^j~.
\label{gapeq}
\ee
This completes the definitions of all the quantities which are used to
describe the thermodynamic potential in Eq.(\ref{thpot}).
%

For numerical evaluations we  choose the parameters as in Ref.\cite{Frank} i.e. 
we write $G=(1-\alpha)G_0$ and $K/2=\alpha G_0$.
The parameter $\alpha$ controls the strength of the instanton interaction while the
value of the quark condensate is determined by the combination of 
parameters : $m=6$ MeV, the three momentum cut off $\Lambda=590$ MeV and the dimensionless coupling $G_0\Lambda^2=2.435$.
These values lead to pion mass in vacuum as 140.2 MeV, pion decay constant of 92.6 MeV and quark condensate 
$\langle\bar u u\rangle=\langle\bar d d\rangle=(-241.5 MeV)^3$, all in reasonable agreement with the experimental values.
This also leads to a vacuum constituent quark mass of 400 MeV. Further, in all these calculations we have taken $\alpha=0.15$
as a reasonable  value interpolated from $\eta-\eta^\prime$ splitting within 3-flavor NJL model \cite{Frank}.

\begin{figure}[h]
\begin{center}
\includegraphics[scale=0.35 ]{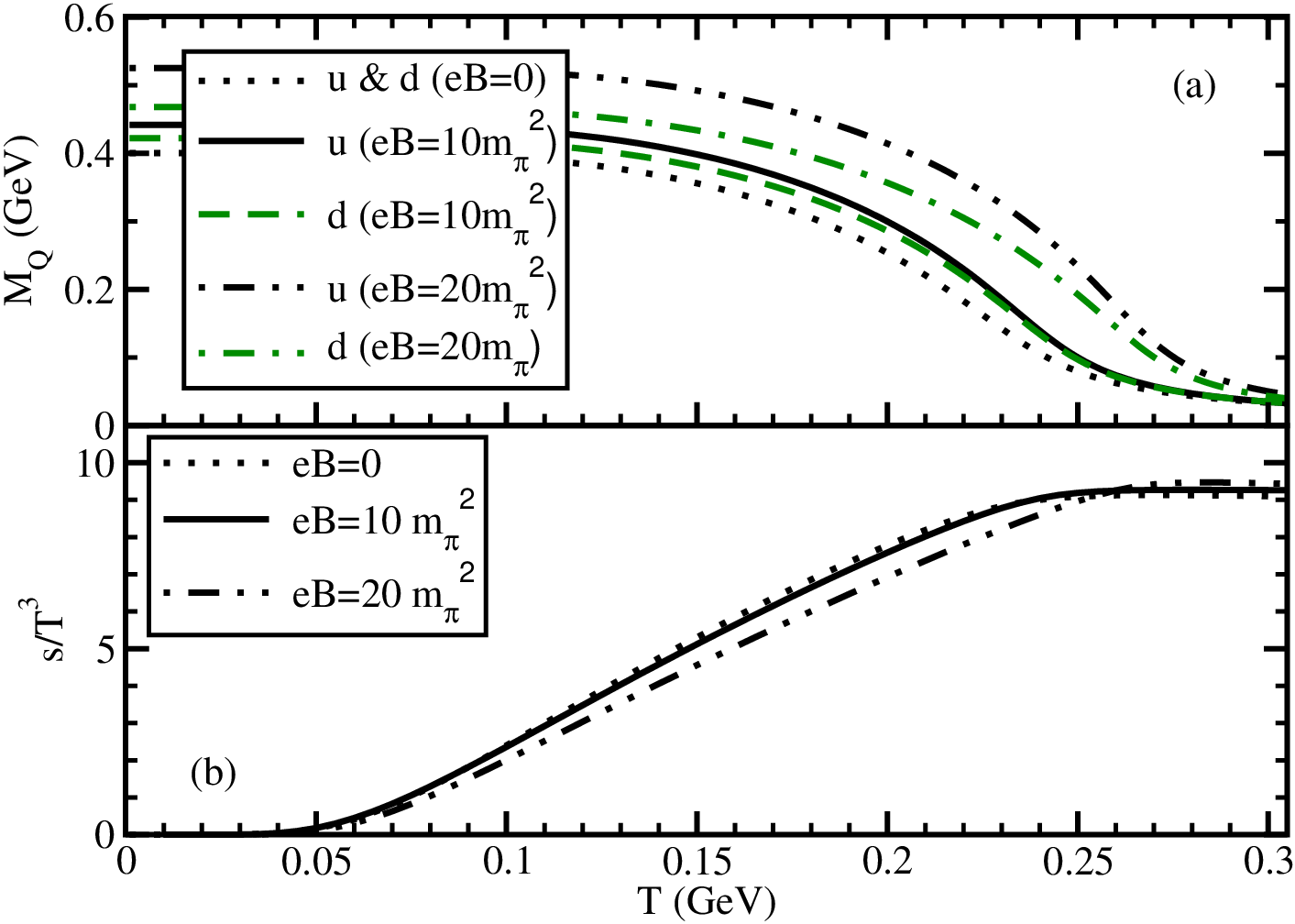}
\caption{$T$ dependence of constituent quark masses $(M_Q)$ and normalized entropy density $(s/T^3)$ for different values of magnetic fields.} 
\label{M_T2}
\end{center}
\end{figure}

Fig.~\ref{M_T2}(a) shows the constituent quark mass as a function
of temperature for different values of magnetic fields.
At $eB=0$, masses of u and d quarks exactly coincide (dotted line), while for 
non-zero $eB$, they split due to different electrical charges of the two quark flavors
and their splitting increases with the magnetic field. Our results reveal the 
magnetic catalysis in entire temperature range and therefore, transition temperature
$T_c$ increases with $B$.
Using this $M_Q(T,eB)$,
one can calculate entropy density $s$ with the help of a quasi-particle relation:
\bea
s&=&\frac{N_c}{\pi^2T}\sum_{i=u,d}\sum_{n=0}^{n_{\rm max}}\alpha_n|q_i|B
\nn\\
&&~~~~~~\int d\vk_z \Big[\frac{\vk^2_z}{\om_n^i} + \om_n^i \Big] f_0(\om_n^i)~,
\label{sQ}
\eea
where $f_0(\om_n^i)$ is Fermi-Dirac distribution function. 
The temperature dependence of normalized 
entropy density $s/T^3$
for $eB=0$ (dotted line), $10m_\pi^2$ (solid line) and $20m_\pi^2$ (dash-dotted line) are
shown in Fig.~\ref{M_T2}(b). We notice that $s$ decreases as $eB$ increases in lower 
temperature domain but all the curves are merged into its Stefan-Boltzmann (SB) limit 
at high temperature region.

\section{Strong magnetic field case}
\label{sec3}
\subsection{Formalism of shear viscosity in presence of strong magnetic field}
\label{sec3A}
Let us first take a brief recapitulation of relaxation time approximation
(RTA) technique to calculate shear viscosity coefficients of a relativistic 
fluid  in absence of any magnetic field (i.e. $B=0$), which is elaborately
given in Refs.~\cite{Gavin,chakrabortty}.
Then, we will come to its corresponding formalism 
in presence of the {\bf strong} magnetic field, well described in
Refs.~\cite{Tuchin_shear,Landau}.

Total energy-momentum tensor of relativistic fluid,
$T^{\mu\nu}=T_0^{\mu\nu} + T_D^{\mu\nu}$ contains
ideal part $T_0^{\mu\nu}=-Pg^{\mu\nu} + (P + \ep)u^\mu u^\nu$
and dissipation part $T_D^{\mu\nu}=\eta U^{\mu\nu}$ (only shear dissipation),
where $P$, $\epsilon$, $u^\mu$ are respectively pressure, energy density and four 
velocity of the fluid. The tensor structure $U^{\mu\nu}$, linked with 
shear viscosity $\eta$, has a form~\cite{chakrabortty}:
\bea
&&U^{\mu\nu}=D^\mu u^\nu + D^\nu u^\mu +\frac{2}{3}
\Delta^{\mu\nu}\partial_\sigma u^\sigma~~~\mbox{with}
\nn\\
&&D^\mu=\partial^\mu - u^\mu u^\sigma \partial_\sigma,~
\Delta^{\mu\nu}=u^\mu u^\nu - g^{\mu\nu}~.
\eea

Now, in terms of four momentum $k^\mu=(\om, \vk)$ 
and thermal distribution function $f_0=1/\{e^{\beta\om} + 1\}$ of quark 
at temperature $T=1/\beta$, one can express the total energy-momentum tensor as
\be
T^{\mu\nu}=\int \frac{d^3\vk}{(2\pi)^3}\frac{k^\mu k^\nu}{\om}\{f_0 + \phi f_0(1 - f_0)\}~,
\label{T_tot}
\ee
%
%
where, the second term in the curly bracket involving the function $\phi$ describes the 
non-equilibrium part for which one can construct the
shear dissipative part $T^{\mu\nu}_D$ of the energy momentum tensor \cite{chakrabortty}. 
In terms of velocity gradient tensor $U^{\mu\nu}$,
the function $\phi$ can be written as $\phi=C k_\mu k_\nu U^{\mu\nu}$.
The unknown $C$ can be obtained as $C=\frac{\tau_c\beta}{2\om}$ by using the 
relativistic Boltzmann equation (RBE), where $\tau_c$ is the relaxation time of the quark
in the medium. 
comparing the coefficients of $U^{\mu\nu}$ from the dissipative part of the energy-momentum
tensor, we finally obtain the expression of shear viscosity coefficient as
%
\be
\eta=\frac{g\beta}{15}\int \frac{d^3\vk}{(2\pi)^3}\frac{\vk^4}{\om^2}\tau_c f_0(1 - f_0)~,
\label{eta_B0}
\ee
where $g=2\times 2\times 2\times 3$ is an  additional input that takes care of the degeneracy factor for 2 flavor 
(isospin symmetric) quark matter.

Now, let us discuss the shear viscosity of the medium in presence of external  magnetic
field, which is decomposed into five independent components. Therefore, the dissipative
part of energy-momentum tensor (in three vector notation) is written as

Now, let us discuss the effect of external magnetic field on the shear viscosity of the medium. 
In presence of a constant background magnetic field, 
the medium can possess five independent components of shear viscosity and the dissipative
part of the  energy-momentum tensor (in three vector notation) can be  written as~\cite{Tuchin_shear,Landau}
\be
T^{ij}_D=\sum^4_{n=0} \eta_n V^{ij}_n=\int \frac{d^3\vk}{(2\pi)^3}\frac{k^i k^j}{\om}\delta f~,
\ee
where
\be
\delta f=\phi f_0(1 - f_0)=\sum^4_{n=0} C_n k^i k^j V^{ij}_n f_0(1 - f_0)
\label{delta_fB}
\ee
and
\be
\phi=\sum^4_{n=0} C_n k^i k^j V^{ij}_n
\label{delta_fB}
\ee
is assumed in terms of same tensorial components $V^{ij}_n$.

Among these 5 components, 4 components ($n=1,..,4$) will be  our
matter of interest as only these components depend on magnetic field, while $n=0$
component remain unaffected by magnetic field. This $n=0$ viscosity component can be
compared with the electrical/thermal conductivity along the direction of magnetic
field, as discussed Refs.~\cite{Sedarkian_cond,Landau}, where they also remain
undisturbed by the external magnetic field.
Hence, ignoring the $\eta_0$ or $V_0^{ij}$ component~\cite{Tuchin_shear,Landau},
one can obtain four shear viscosity coefficients as
\be
\eta^i_{(n=1,2,3,4)}=\frac{2g_i}{15}\int \frac{d^3\vk}{(2\pi)^3}\frac{\vk^4}{\om^i}C^i_{(n=1,2,3,4)} f^i_0(1- f^i_0)~,
\label{eta_CB}
\ee
where the unknown $C^i_n$ again will be determined with the help of  
the RBE but in two step approximations.
Since the magnetic field will destroy the degeneracy of $u$ and $d$ quark masses,
therefore, energy $\om^i$, distribution function $f^i_0$ and $C^i_n$ in Eq.~(\ref{eta_CB})
carry the flavor index $i$. The $g_i=2\times 2\times 3$ is degeneracy factor of each flavor.

As a first approximation, the particle relaxation time $\tau_c$ in the RBE is ignored by assuming that 
the deviation from equilibrium due to the strong magnetic field is much
larger than that due to the particle collisions.
Therefore, we get a magnetic field induced relaxation time $\tau^i_B=1/\om^i_B$, where
\be
\om^i_B={q_i}B/\om^i~,~({q_i}=+\frac{2}{3}e,~-\frac{1}{3}e~{\rm for}~i=u,~d)
\label{om_B}
\ee
is the synchrotron frequency of quark. So the strong field limit will be
established if we can show that  $\tau^i_c>>\tau_B^i$.
As a  first approximation of RBE~\cite{Tuchin_shear,Landau}, 
we ignore $\tau_c^i$ leading to the the coefficients $C^i$ getting related to the field induced relaxation time $\tau_B$ as
\be
C^i_1=C^i_2=0~,~{\rm and}~C^i_4=2C^i_3=\frac{\tau^i_B\beta}{2\om^i}~.
\label{c1c20}
\ee
Now, in a second approximation,
a collisional or thermal width 
$\Gamma^i_c=1/\tau^i_c$, obeying the inequality $\Gamma^i_c << \om^i_B$ or $\tau^i_c >> \tau^i_B$,
is considered which leads to the relation~\cite{Tuchin_shear}:
\be
C^i_2=4 C^i_1= \frac{\Gamma^i_c}{\om^i_B}C^i_4=\frac{\Gamma_c^i}{2\om^i_B}C^i_3~,
\ee
with $C^i_4=2C^i_3=\frac{\tau^i_B\beta}{2\om^i}$. 
Thus, in presence of constant background magnetic field $B$, the expressions of the four components
of the shear viscosity for $i=u/d$ quark are
\bea
\eta^i_2=4\eta^i_1&=&\frac{g_i\beta}{15}\int \frac{d^3\vk}{(2\pi)^3}[f^i_0\{1- f^i_0\}]
\nn\\
&&~~~~~~\left\{\left(\frac{\Gamma^i_c}{\om^i_B}\right)\left(\frac{1}{\om^i_B}\right)\right\}
\left(\frac{\vk^2}{\om^i}\right)^2,
\label{eta_2}
\eea
and
\bea
&&\eta^i_4=2\eta^i_3
\nn\\
&=&\frac{g_i\beta}{15}\int \frac{d^3\vk}{(2\pi)^3}
[f^i_0\{1- f^i_0\}]\left\{\frac{1}{\om^i_B}\right\}
\left(\frac{\vk^2}{\om^i}\right)^2~.
\label{eta_4}
\eea
If we compare Eqs.~(\ref{eta_2}), (\ref{eta_4}) with Eq.~(\ref{eta_B0}), then
we can get a physical interpretation of these shear viscosity components.
In the perpendicular plane to the external magnetic field,
the momentum transfer due to shear stress is independent of the particle collisions and  
 will be proportional to the field induced relaxation ($\tau_B=1/\om_B$) which is basically the inverse of  
the synchrotron frequency. In other words, rotational motion of the charged particles
with corresponding synchrotron frequency provides the required momentum transfer for generating
shear stress along the tangential directions, located in the perpendicular plane with respect to 
the magnetic field. This strength
of shear stress, velocity gradient and its proportional coefficients $\eta_3$,
$\eta_4$ are completely originated due to (strong) magnetic field background.

In other possible tangential directions, 
both the collisional and rotational energies take part in momentum transfer.
Therefore, the fraction $\Gamma_c/\om_B$ is required for fixing the proportional strength
of viscosities $\eta_1$ and $\eta_2$. The corresponding
relaxation time for these components becomes 
$\Big[\big(\frac{\Gamma_c}{\om_B}\big)\frac{1}{\om_B}\Big]^{-1}$.

\subsection{Results of strong field case}
\label{sec3B}
 For the strong field limit to be valid 
$\tau_c$ should be much largerr than 
$\tau_B=1/\om_B=\om^k_Q/e_QB=\{\vk^2 +m_Q^2\}^{1/2}\Big/e_QB$~\cite{Tuchin_shear},
which is the inverse of the synchrotron frequency $\om_B$.
The assumption $\tau_c>>\tau_B$ is the basis of strong field case formalism,
discussed in earlier Sec.~(\ref{sec3A}).
After calculating $\tau_c$ microscopically, one will be  able to know  
whether the value of $\tau_c$ satisfy $\tau_c>>\tau_B$ or strong field limit or not.
\begin{figure} 
\begin{center}
\includegraphics[scale=0.35]{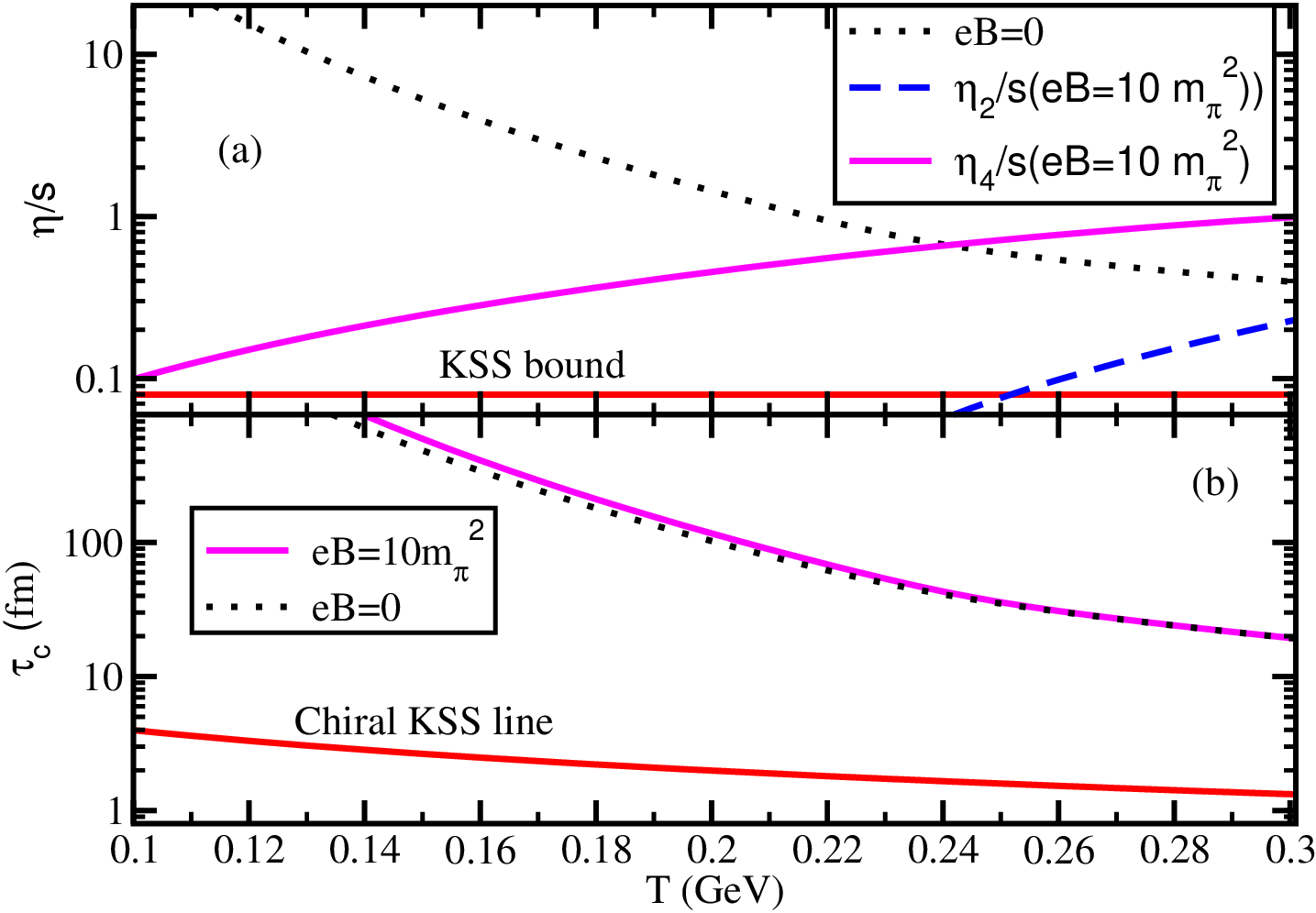}
\caption{(a) $T$ dependence of $\eta/s$ at $eB=0$ (dotted line) and $\eta_2/s$ (blue dash line),
$\eta_4/s$ (pink solid line) at $eB=10 m_\pi^2$.
(b) $T$ dependence of $\tau_c$ at  $eB=0$ (dotted line) and $eB=10 m_\pi^2$ (pink solid line).} 
\label{es_Ttau}
\end{center}
\end{figure}
Here, we will attempt to obtain $\tau_c(T,eB)$ in a  microscopic calculation.
For this purpose, let us start with $B=0$ case with standard expression of collisional 
relaxation time $\tau_c$ or thermal width,
\bea
&&\Gamma_c(T,\vk_a)=\frac{1}{\tau_c}
\nn\\
&&=\sum_{b}\int \frac{d^3k_b}{(2\pi)^3}\sigma_{ab}(T,\vk_a,\vk_b) v_{ab}(T,\vk_a,\vk_b) 
f_b(T,\vk_b)~,
\nn\\
\label{G_KTmu}
\eea
where 
\be
v_{ab}(T,\vk_a,\vk_b)=\frac{\{(\om_a +\om_b)^2-4M_Q^2(T)\}^{1/2}(\om_a +\om_b)}{2\om_a\om_b}
\ee
is relative velocity with $\om_{a,b}=\{\vk_{a,b}^2+M_Q^2(T)\}^{1/2}$. 
To map grossly the scattering strength of NJL dynamics, let us calculate cross section
$\sigma_{ab}$ from simple four quark contact diagram, shown inside Fig.~\ref{es_Ttau}(b).
To do it, we use the standard quantum field theoretical relation of $2\rightarrow 2$ scattering,
\be
\sigma_{ab}=\frac{1}{16\pi s}{\overline{|M|_{ab}^2}}~,
\ee
where $s=(\om_a+\om_b)^2$ and 
\bea
{\overline{|M|_{ab}^2}}&=&\frac{1}{2\times 2}G^2 16\Big(\frac{s}{2}\Big)^2
\nn\\
&=&G^2s^2~,~s=(\om_a +\om_b)^2~.
\eea
Hence, we get a temperature and momentum dependence cross section 
$\sigma_{ab}(T,\vk_a,\vk_b)=\frac{G^2}{16\pi}s(T,\vk_a,\vk_b)$.

By maintaining electric charge conservation, we will get 12 possible $2\rightarrow 2$ 
$(ab\rightarrow a'b')$ scattering processes : 
\bea
&&u{\bar u}\rightarrow u{\bar u},~ u{\bar d}\rightarrow u{\bar d},~ u{\bar u}\rightarrow d{\bar d},~ uu\rightarrow uu,~ 
\nn\\
&&ud\rightarrow ud,~ {\bar u}{\bar u}\rightarrow {\bar u}{\bar u},~  {\bar u}{\bar d}\rightarrow {\bar u}{\bar d},~  d{\bar d}\rightarrow d{\bar d},~ 
\nn\\
&& d{\bar d}\rightarrow u{\bar u},~  
 d{\bar u}\rightarrow d{\bar u},~ dd\rightarrow dd,~ {\bar d}{\bar d}\rightarrow {\bar d}{\bar d},~
\eea
So fixing any initial particle $a$ as probe particle, we have to take summation of $b$ to calculate $\Gamma_c(T,\vk_a)$
Taking momentum average of probe particle, we get only $T$ dependent quark width,
\bea
&&\Gamma_{c}(T)=\frac{1}{\tau_{c}}
\nn\\
&&=\frac{\int \frac{d^3k_a}{(2\pi)^3}\Gamma(T,\vk_a) f_a(T,\vk_a)} 
{\int\frac{d^3k_a}{(2\pi)^3} f_a(T,\vk_a)}~.
\nn\\
\label{GQ_T}
\eea
So we find that temperature dependent mainly coming from thermodynamical phase space
and $M_Q(T)$. If we go for simplified extension of finite magnetic field picture by
replacing $M_Q(T,eB)$ in Eq.~(\ref{GQ_T}), we can get $\Gamma_c(T,eB)$.
Fig.~\ref{es_Ttau}(b) shows
the $\tau_c(T, eB)$ at $eB=0$ (black dotted line), $eB=10 m_\pi^2$ (pink solid line). 

Due to increase in the number density of the particles in the medium with temperature, the
collisional frequency increases and relaxation time decreases
with $T$. The $eB$ dependence of $\tau_c$ enters via $eB$ dependence of constituent
quark mass $M_Q(eB)$. Increasing function $M_Q(eB)$ can suppress the number density,
which make $\Gamma_c$ decrease and $\tau_c$ increase with $eB$. Being proportional
to the decreasing function $\tau_c(T)$ for $eB=0$, $\eta/s$ decreases with $T$. 

Let us note that in the behaviour of $\eta /s$ with temperature arises from two competing quatities that depend upon
temperature. Due to thermodynamic phase space factor, this ratio is an increasing function of tempearture while
the relaxation time decrease with temperature. For constant relaxation time, due to thermal phase space factor, $\eta /s$
increase with temperature. This can be easily found out for mass less ideal gas behaviour of the expression for $\eta/s$.
On the otherhand, for  the decreasing behaviour of temperature dependent $\tau$, dominates over the increasing behaviour 
arising from the thermal phase space making the ratio decreasing with temperature,
as may be noticed in dotted line of 
Fig.~\ref{es_Ttau}(a). 

Another noticeable thing is that for contact diagram of 
$2\rightarrow 2$ scattering processes, NJL model estimate is rather large for  $\tau_c$,
which is quite far from chiral KSS line $\tau_c(T)=\frac{5}{4\pi T}$, shown
by red solid line in Fig.~\ref{es_Ttau}(b). The chiral KSS line comes from the 
demand of $\frac{\eta}{s}=\frac{1}{4\pi}$ for massless particle. Therefore, in Fig.~\ref{es_Ttau}(a),
black dotted line is also quite far from red horizontal line, denoted KSS value 
$\frac{\eta}{s}=\frac{1}{4\pi}$. 

For this high value of $\tau_c$, $eB=10 m_\pi^2$
can be considered as strong field limit case because
$\tau_B$ remain within the range 0.8-3 fm. So we can safely say that
at $eB=10 m_\pi^2$, we can consider $\tau_c>> \tau_B$ or strong field
limit case~\cite{Tuchin_shear}.
It is interesting to notice in Eqs.~(\ref{eta_2}) that the position of
$\tau_c$ for strong field case becomes inverse ($\eta_2\propto 1/\tau_c$), therefore, $\eta_2/s$ becomes
an increasing function of $T$, as shown by pink solid line in Fig.~\ref{es_Ttau}(a). 
%
%
%
Hence, in strong field limit, $\eta_{1,2}\propto 1/\tau_c$ follow 
opposite trend with respect without field case $\eta\propto\tau_c$.
When we come to the Hall-type viscosity $\eta_{3,4}\propto\tau_B$, 
which is appeared as dissipation-free completely as it becomes independent
of $\tau_c$. $\eta_{3,4}$  increases with $T$ because of its phase space part,
which cane be realized from $\eta_4/s$ curve (blue dash line) in Fig.~\ref{es_Ttau}(a).

Now, for simplest contact diagram calculation, we are getting very large value of $\tau_c$
but it can not be expected in RHIC or LHC matter, whose life time is approximately 10 fm.
So, the strong field case can't be applicable for RHIC or LHC matter, whose $\tau_c$ is expected
to be small, at least smaller than 10 fm. We might find alternative possible
diagrams, which can provide small $\tau_c$. Refs.~\cite{hm1,hm2,hm3,SG_NJL1,SG_NJL2} have 
obtained very small $\tau_c$, relevant to RHIC or LHC matter through meson exchange type
diagram, whose calculation in presence of magnetic field is not at all very straight forward.
It might be considered as future challenging topics. Instead of calculating
smaller $\tau_c(T)$, we can take it as parameter and examine the impact of its smaller value.
When we consider small value of $\tau_c$ ($<10$ fm) at $eB=10 m_\pi^2$, 
the inequality $\tau_c>>\tau_B$ does not hold. So instead of considering 
strong field limit, we might have to find some general structure of $\eta_n$,
which has been attempted in next section.

\section{From strong field to moderate fields}
\label{sec4}
\subsection{Modified formalism of shear viscosity}
\label{sec4A}
In this section,
we will attempt to find a guess general structure of shear viscosities, which can be applicable
for any value of $\tau_B$ and $\tau_c$.

We have found that the $\tau_c$ in Eq.~(\ref{eta_B0}) for $B=0$ is basically
replaced by effective relaxation time $\tau^{\rm eff}_{1,2}=\frac{\tau_B^2}{\tau_c}$ for $\eta_{1,2}$
and $\tau^{\rm eff}_{3,4}=\tau_B$ for $\eta_{3,4}$. Let us guess an ansatz  of 
effective relaxations:
\bea
\tau^{\rm eff}_{1}&=&\tau_c
\frac{1}{4\{\frac{1}{4}+(\tau_c/\tau_B)^2\}}
\nn\\
\tau^{\rm eff}_{2}&=&\tau_c
\frac{1}{\{1+(\tau_c/\tau_B)^2\}}
\nn\\
\tau^{\rm eff}_{3}&=&\tau_c
\frac{\tau_c/\tau_B}{2\{\frac{1}{4}+(\tau_c/\tau_B)^2\}}
\nn\\
\tau^{\rm eff}_{4}&=&\tau_c
\frac{\tau_c/\tau_B}{\{1+(\tau_c/\tau_B)^2\}}
\label{t_gen}
\eea 
which might be consider as their general structure, because
in the limit of $\tau_c>>\tau_B$, we get
\bea
4\tau^{\rm eff}_{1}&=&\tau^{\rm eff}_{2}=\frac{\tau_B^2}{\tau_c}
\nn\\
2\tau^{\rm eff}_{3}&=&\tau^{\rm eff}_{4}=\tau_B~.
\eea
It means that we will get back Eqs.~(\ref{eta_2}) and (\ref{eta_4}) for 
the strong field limit ($\tau_c>>\tau_B$).

Using that general structure of relaxation (\ref{t_gen}) in Eqs.~(\ref{eta_2}) and (\ref{eta_4}) we 
will get general expressions of shear viscosity components:
\bea
\eta_1&=&\frac{g\beta}{15}\int \frac{d^3\vk}{(2\pi)^3}\left(\frac{\vk^2}{\om}\right)^2\tau_c
\frac{1}{4\{\frac{1}{4}+(\tau_c/\tau_B)^2\}}
\nn\\
&&~~~~~~~~~~ [f_0\{1- f_0\}]
\label{eta1_B}
\\
\eta_2&=&\frac{g\beta}{15}\int \frac{d^3\vk}{(2\pi)^3} \left(\frac{\vk^2}{\om}\right)^2\tau_c
\frac{1}{1+(\tau_c/\tau_B)^2}
\nn\\
&&~~~~~~~~~~ [f_0\{1- f_0\}]
\label{eta2_B}
\\
\eta_3&=&\frac{g\beta}{15}\int \frac{d^3\vk}{(2\pi)^3}\left(\frac{\vk^2}{\om}\right)^2 \tau_c
\frac{\tau_c/\tau_B}{2\{\frac{1}{4}+(\tau_c/\tau_B)^2\}}
\nn\\
&&~~~~~~~~~~ [f_0\{1- f_0\}]
\label{eta3_B}
\\
\eta_4&=&\frac{g\beta}{15}\int \frac{d^3\vk}{(2\pi)^3}\left(\frac{\vk^2}{\om}\right)^2 \tau_c
\frac{\tau_c/\tau_B}{1+(\tau_c/\tau_B)^2}
\nn\\
&&~~~~~~~~~~ [f_0\{1- f_0\}]
\label{eta4_B}
\eea

\subsection{Modified results for moderate fields}
\label{sec4B}
\begin{figure} 
\begin{center}
\includegraphics[scale=0.35]{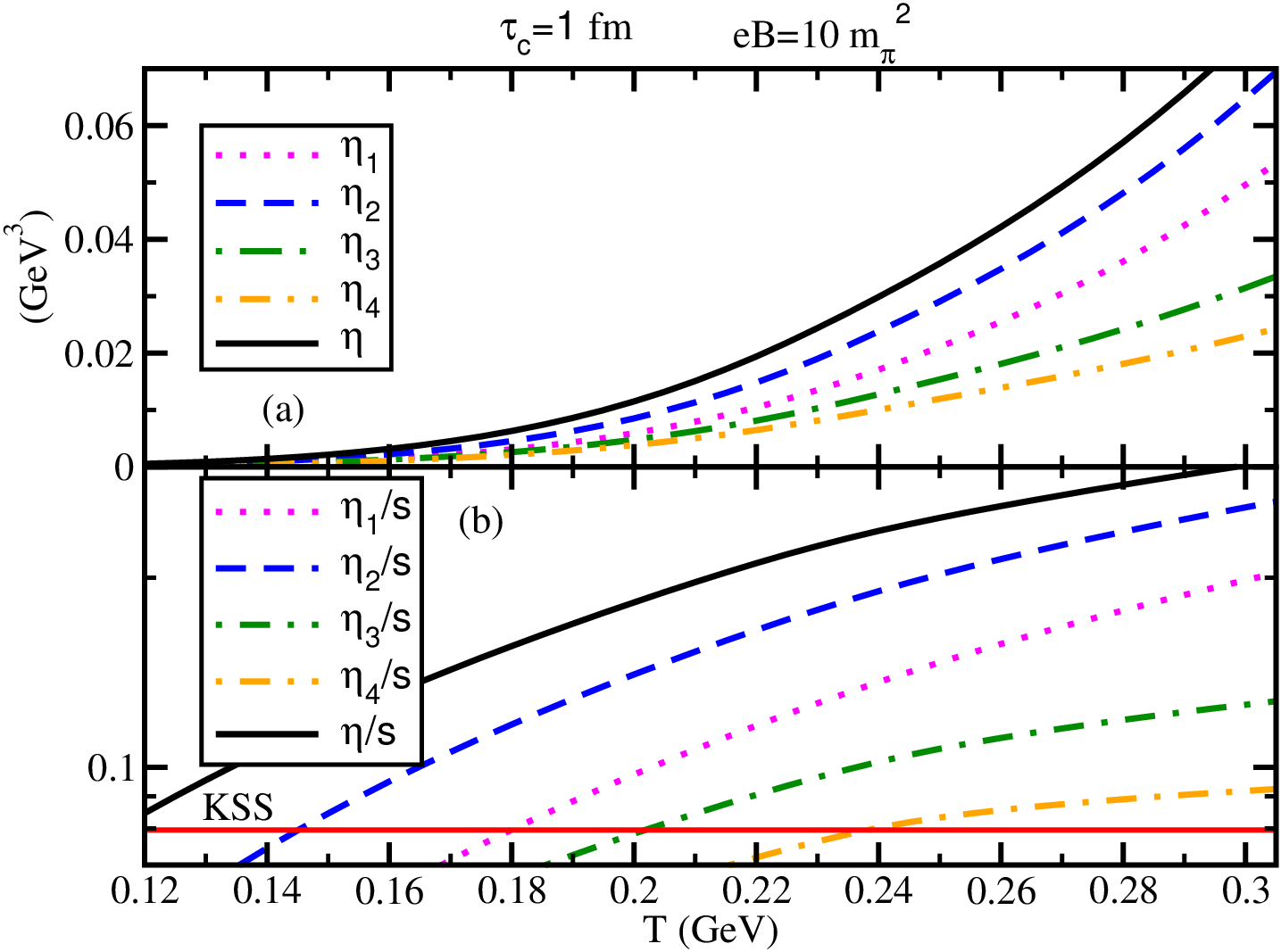}
\caption{(a) $T$ dependence of $\eta$ at $eB=0$ (solid line) and $\eta_{1,2,3,4}$ (dotted, dash, dash-dotted, dash-double-dotted lines) at 
$eB=10 m_\pi^2$. (b) Corresponding viscosity to entropy density ratios.} 
\label{eta1234_T}
\end{center}
\end{figure}
\begin{figure} 
\begin{center}
\includegraphics[scale=0.35]{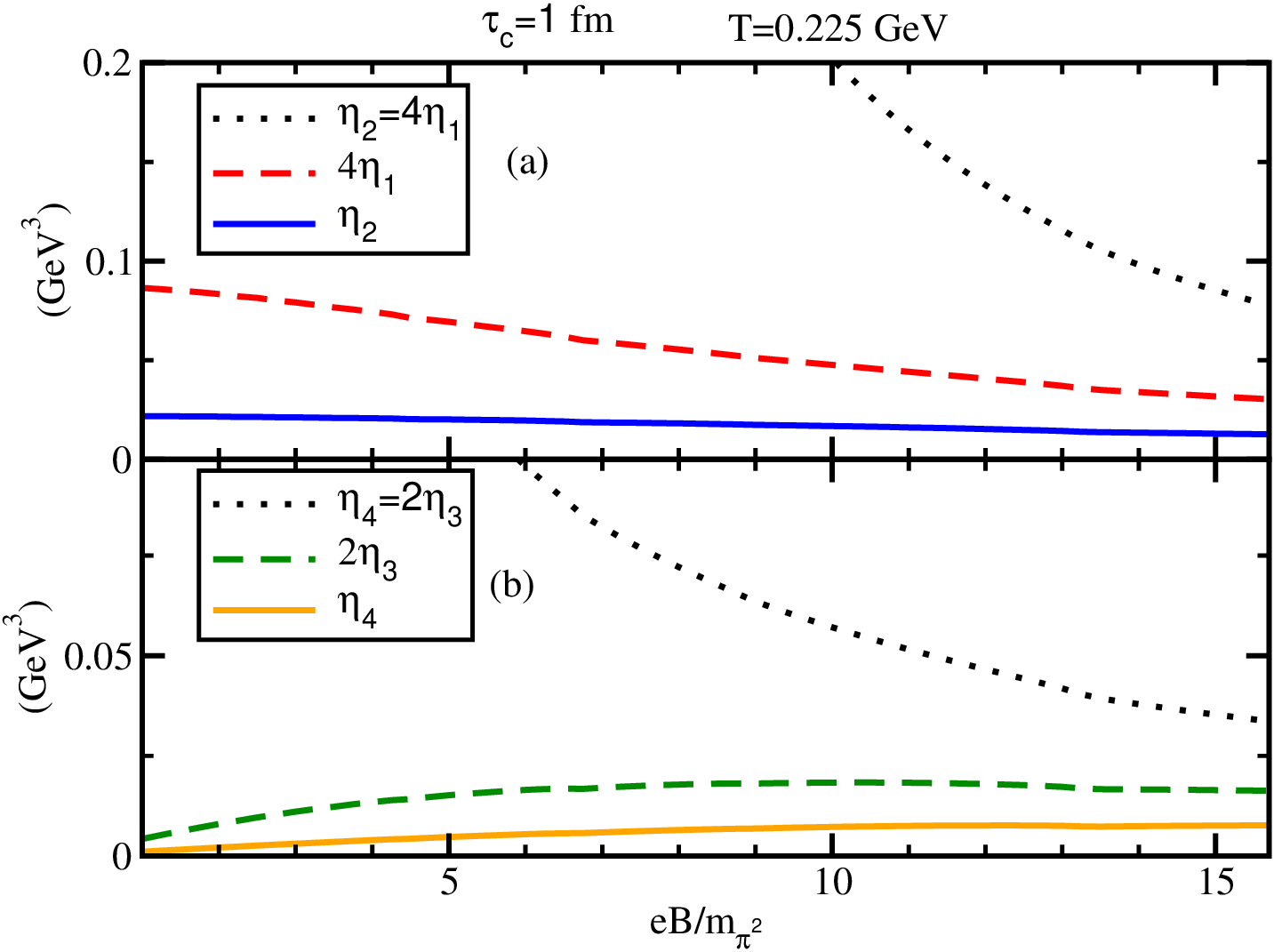}
\includegraphics[scale=0.35]{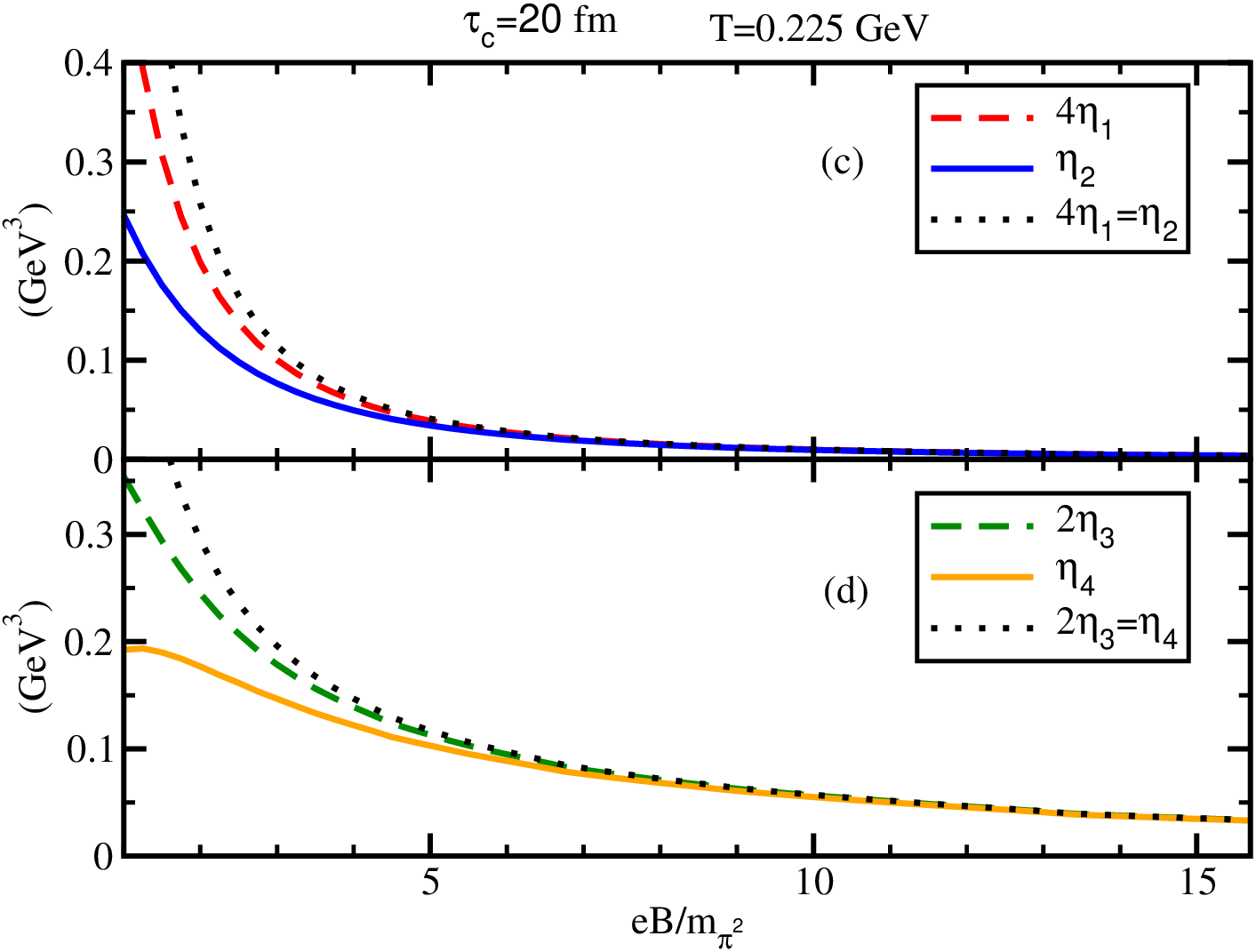}
\caption{For $\tau_c=1$ fm, (a) $4\eta_1$, $\eta_2$ from Eqs.~(\ref{eta1_B}), (\ref{eta2_B}) respectively and its
strong field limit (where $4\eta_1=\eta_2$) from Eq.~(\ref{eta_2}). Similarly, for $\tau_c=1$ fm, 
(b) $2\eta_3$, $\eta_4$ from Eqs.~(\ref{eta3_B}), (\ref{eta4_B}) respectively and its
strong field limit (where $2\eta_3=\eta_4$) from Eq.~(\ref{eta_4}).
(c) and (d) are same as (a) and (b) for $\tau_c=20$ fm.} 
\label{eta1234_B}
\end{center}
\end{figure}
 \begin{figure} 
 \begin{center}
 \includegraphics[scale=0.35]{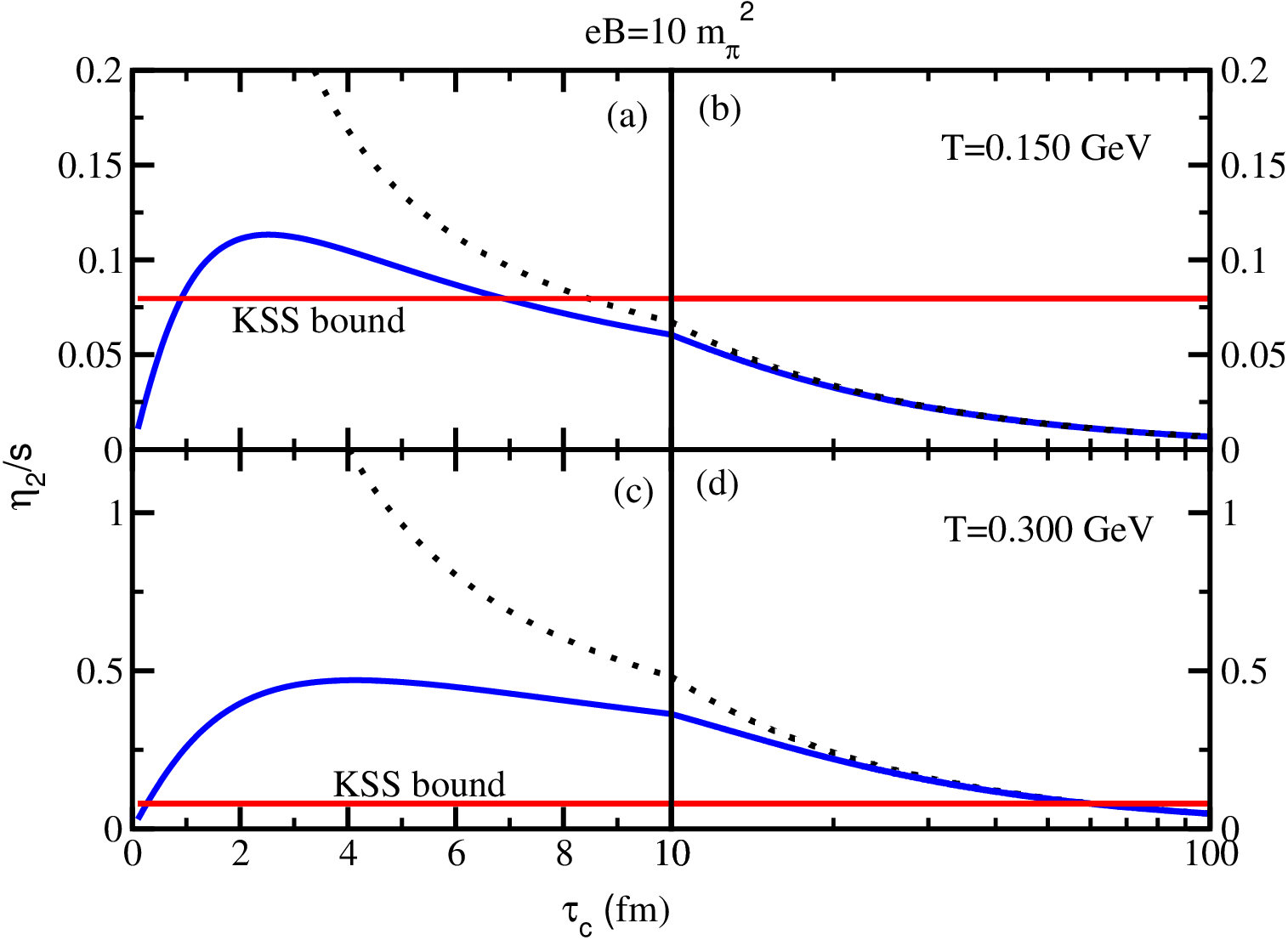}
 \caption{$\tau_c$ dependence of $\eta_2/s$ from Eq.~(\ref{eta2_B}) (solid line) and Eq.~(\ref{eta_2}) 
 (dotted line) at $eB=10m_\pi^2$, $T=0.150$ GeV in the range of 
 (a) $\tau_c < 10$ fm and (b) $\tau_c > 10$ fm; and at $T=0.300$ GeV in the range of 
 (c) $\tau_c < 10$ fm and (d) $\tau_c > 10$ fm. Straight horizontal red line denotes the KSS bound.}
 \label{es24_TB}
 \end{center}
 \end{figure}
Using Eqs.~(\ref{eta1_B}), (\ref{eta2_B}), (\ref{eta3_B}), (\ref{eta4_B}), we have first
plotted $\eta_{1,2,3,4}$ vs $T$ in Fig.~\ref{eta1234_T}(a) and then their normalized values
$\eta_{1,2,3,4}/s$ in Fig.~\ref{eta1234_T}(b). Interesting point is that all components of 
shear viscosity in presence of magnetic field is smaller than its isotropic value in absence
of magnetic field. In Fig.~\ref{eta1234_T}(b), we notice that KSS line is crossing
different curves at different temperature. Here, we find that at fixed $\tau_c$, perfect
fluid nature will be developed in quark matter at higher temperature for $B\neq 0$ 
with respect to its $B=0$ case. It reflects that for fixed interaction, magnetic
field push the system towards KSS bound, while temperature kick away from the bound. 
To zoom in the fact, we have plotted $\eta_{1,2,3,4}$ vs $eB/m_\pi^2$ in Fig.~\ref{eta1234_B},
where decreasing trend of $\eta_{1,2}$ with magnetic field is clearly observed. Magnetic
field dependent of Hall-type coefficients $\eta_{3,4}$ behave little different because of
its anisotropic structure $\frac{(\tau_c/\tau_B)}{1+(\tau_c/\tau_B)^2}$, which increases with
$B$ for $\tau_c/\tau_B<1$ but decreases with $B$ for $\tau_c/\tau_B>1$. Therefore, 
we get increasing $\eta_{3,4}(B)$ for $\tau_c=1$ fm and decreasing $\eta_{3,4}(B)$ 
for $\tau_c=20$ fm, as displayed in Fig.~\ref{eta1234_B}(b) and (d).

Another interesting point has also been shown in Fig.~\ref{eta1234_B}. It is
regarding the merging of general anisotropic shear viscosity components with their
strong field limit estimation. At strong field limit, $\eta_2=4\eta_1$
and $\eta_4=2\eta_3$, whose expressions are given in Eqs.~(\ref{eta_2}), (\ref{eta_4}).
These strong field limit estimations of $\eta_2=4\eta_1$ and $\eta_4=2\eta_3$ curves
are plotted by dotted line in Fig.~\ref{eta1234_B}(a) and (b) respectively for $\tau_c=1$ fm.
On the other hand, general anisotropic components of shear viscosity can be obtained from
Eqs.~(\ref{eta1_B}), (\ref{eta2_B}), (\ref{eta3_B}), (\ref{eta4_B}) and plotted $4\eta_1$ (red dash line),
$\eta_2$ (blue solid line), $2\eta_3$ (green dash line), $\eta_4$ (orange solid line) 
in Fig.~\ref{eta1234_B}(a) and (b) for $\tau_c=1$ fm. We notice that these curves are not 
merging in any point of $B$-axis up to $eB=15m_\pi^2$ but 
when we use $\tau_c=20$ fm in Figs.~\ref{eta1234_B}(c,d), they are merging after  $eB=6m_\pi^2$.
It tells that strong field limit might be good approximation for $\tau_c> 10$ fm but for RHIC
or LHC matter, whose $\tau_c< 10$ fm, strong field limit expressions might not be considered
as a good approximated estimation. This picture will be more clear in Fig.~\ref{es24_TB}, which
exposes the $\tau_c$ dependence of $\eta_2/s$ from Eq.~(\ref{eta2_B}) (solid line) and Eq.~(\ref{eta_2}) 
 (dotted line) at $eB=10m_\pi^2$, $T=0.150$ GeV in the range of 
 (a) $\tau_c < 10$ fm and (b) $\tau_c > 10$ fm. Here, we find how general $\eta_2/s$
 is merging with its strong field limit curves in second zone ($\tau_c > 10$ fm), while
 they are quite far in first zone ($\tau_c<10$ fm). Same qualitative pattern is also noticed
 for $T=0.300$ GeV in Fig.~\ref{es24_TB}(c,d). From the crossing of KSS line in Fig.~\ref{es24_TB},
 we can say that nearly perfect fluid nature can be obtained for two different values of $\tau_c$ within $\tau_c<10$ fm zone at 
 $eB=10m_\pi^2$, $T=0.150$ GeV. We can denote them as $\tau^{\mp}_c$. The values of $\tau_c^-$ is seen below 1 fm, so it might
 not be reached in RHIC or LHC matter. So $\tau_c^+$ is more phenomenological point.
 Interestingly, we notice that $\tau_c^+$ remain $\tau_c>10$ fm zone for $T=0.300$ GeV.
 Hence, alternatively we can say that to build nearly perfect fluid nature in
 higher temperature quark matter, we need higher magnetic field if we want it for $\tau_c<10$ fm zone.
 It is quite possible for RHIC or LHC matter, having $\tau_c<10$ fm zone, where high magnetic
 field decays with time and temperature. I means that as we approach higher temperature, that matter
 can face higher magnetic field in experiments. So there might be a compensating role of
 temperature and magnetic field to build nearly perfect fluid nature in RHIC or LHC matter.


%

%

\section{Summary} 
\label{sec6}
We have studied shear viscosity of quark matter 
in a uniform magnetic field background, where the medium looses its
isotropic property. 
Due to this anisotropic nature, one can get more than one components
of shear viscosity, denoted by $\eta_1$, $\eta_2$, $\eta_3$ and $\eta_4$, 
which are ultimately reduced to two main components in strong field limit through 
relations $4\eta_1=\eta_2$
and $2\eta_3=\eta_4$. 
We know that isotropic shear viscosity $\eta$ in absence of magnetic field
is mainly governed by two parts - the phase space and the relaxation time.
Here also $\eta_2$ and $\eta_4$ can be casted into the similar structure with
phase space and relaxation time parts.
The relaxation time of $\eta_4$ is inversely proportional to synchrotron frequency $\om_B$
and relaxation time of $\eta_2$ is $\frac{\Gamma_c}{(\om_B)^2}$ in strong field limit, where
collisional thermal width $\Gamma_c$ of medium constituents will be much smaller that its
synchrotron frequency i.e. $\Gamma_c<<\om_B$. Although, a large values of $\Gamma_c$
is expected for strongly-coupled RHIC or LHC matter. To describe that zone, we need a
general structure of $\eta_{1,2,3,4}$, which don't follow the relations $4\eta_1=\eta_2$
and $2\eta_3=\eta_4$ below the strong field domain.

We have used
the formalism of NJL model in presence of magnetic field to describe the 
magneto-thermodynamics of quark matter and we get a temperature and magnetic field
dependent quark mass, which will enter to the phase space factors of $\eta_{1,2,3,4}$.
In strong field limits, all components decrease with $B$ but in weak field case, the Hall-type
viscosities $\eta_{3,4}$ increase with $B$.
Along with the constant value $\tau_c$, we have also calculated $T$ dependent of $\tau_c$
from simple contact diagram of $2\rightarrow 2$ scattering processes, 
coming from the interaction Lagrangian density of NJL model. Replacing $T$ dependent quark mass 
$M_Q(T)$ by $T$, $eB$ dependent quark mass $M_Q(T,eB)$, we have extended the expression
of relaxation time from $\tau_c(T)$ to $\tau_c(T,eB)$.
Scattering probability ($\Gamma_c$) proportionally increases with density of medium,
which increases with $T$ due to statistical reason and decreases with $eB$ due to mass enhancement. 
Hence, relaxation time $\tau_c(T,eB)$ decreases with $T$ and increases with $eB$.
In absence of magnetic field, shear viscosity to entropy density ratio decreases with $T$ as it
is proportional to relaxation time but it increases with $T$ in strong field picture as it is inversely
proportional to the relaxation time. So, transition from without to with magnetic field
picture, $T$ dependence of viscosity to entropy density ratio transforms from decreasing to
increasing trends. 

In present work, we have calculated $\tau_c(T,eB)$ from simplest contact diagram $2\rightarrow 2$
scattering processes, which provide a large $\tau_c$, where $eB=10m_\pi^2$ can safely be considered as
strong field case. However, to describe RHIC or LHC matter with small $\tau_c$, we have to consider
the general structure of $\eta_{1,2,3,4}$ and better interaction picture, which can map strongly-coupled
matter. We keep this problem as our future goal, which definitely provide an up-gradation of research on transport
properties of quark matter under external magnetic field. 
{\bf Acknowledgment:} 
SG acknowledges to Indian Institute of Technology (IIT) Bhilai, funded by 
Ministry of Human Resource Development (MHRD) as well as earlier D. S. Kothari
fellowship, University Grants Commission (UGC) under grant No. F.4-2/2006
(BSR)/PH/15-16/0060. 
SG and AM thank to Pracheta Singha for initial constructive discussion
on this work.


\begin{thebibliography}{99}
%
\bibitem{Schafer_Rev}
T. Schafer, D. Teaney,
Rep. Prog. Phys. 72 (2009) 126001.
%
\bibitem{Tuchin_Rev} K.~Tuchin, 
Adv. High Energy Phys. 2013 (2013). 
%
\bibitem{Skokov:2009qp} 
V.~Skokov, A.~Y.~Illarionov and V.~Toneev,
Int.\ J.\ Mod.\ Phys.\ A {\bf 24}, 5925 (2009).
%
\bibitem{Voronyuk:2011jd} 
V.~Voronyuk, V.~D.~Toneev, W.~Cassing, E.~L.~Bratkovskaya, V.~P.~Konchakovski and S.~A.~Voloshin,
Phys.\ Rev.\ C {\bf 83}, 054911 (2011).
%
\bibitem{Bzdak:2011yy} 
A.~Bzdak and V.~Skokov,
Phys.\ Lett.\ B {\bf 710}, 171 (2012).
%
\bibitem{Deng:2012pc} 
W.~T.~Deng and X.~G.~Huang,
Phys.\ Rev.\ C {\bf 85}, 044907 (2012).
%
\bibitem{Deng:2014uja} 
W.~T.~Deng and X.~G.~Huang,
Phys.\ Lett.\ B {\bf 742}, 296 (2015).
%
\bibitem{jens_rmp}
J.~O.~Andersen, W.~R.~Naylor and A.~Tranberg,
Rev. Mod. Phys. {\bf 88}, 025001 (2016).
%
\bibitem{Miransky}
V. A. Miransky and I. A. Shovkovy, 
Phys. Rept. {\bf 576},1 (2015). 
%
\bibitem{Marco1} R. Gatto, M. Ruggieri,
Phys. Rev. {\bf D 83} (2011) 034016.
%
\bibitem{Marco2} R. Gatto, M. Ruggieri,
Lect.Notes Phys. 871 (2013) 87; arXiv:1207.3190 [hep-ph].
%
\bibitem{Boomsma:2009yk} 
J.~K.~Boomsma and D.~Boer,
Phys.\ Rev.\ D {\bf 81}, 074005 (2010).
%
\bibitem{BC_HM1}B. Chatterjee, H. Mishra, A. Mishra,
Phys.\ Rev.\ {\bf D 84}, 014016 (2011).
%
\bibitem{BC_HM2}B. Chatterjee, H. Mishra, A. Mishra,
Phys.\ Rev.\ {\bf D 91}, 034031 (2015).
%
\bibitem{Bali} 
G.~S.~Bali, F.~Bruckmann, G.~Endrodi, Z.~Fodor, S.~D.~Katz, S.~Krieg, A.~Schafer and K.~K.~Szabo,
J. High Energy Phys. {\bf 1202}, 044 (2012).
%
\bibitem{Bornyakov}
V.~G.~Bornyakov, P.~V.~Buividovich, N.~Cundy, O.~A.~Kochetkov and A.~Schafer,
Phys. Rev. D {\bf 90}, 034501 (2014).
%
\bibitem{Ayala_LSM1} 
A. Ayala, M. Loewe,  A. Z. Mizher and Zamora, R.,
Phys. Rev.D {\bf 90}, 036001  (2014).
%
\bibitem{Ayala_LSM2} 
A.~Ayala, M.~Loewe and R.~Zamora,
Phys.\ Rev.\ D {\bf 91}, 016002 (2015)
%
%
\bibitem{Krein1}R.L.S. Farias, K.P. Gomes, G.I. Krein, M.B. Pinto
Phys. Rev. {\bf C 90}, 025203 (2014). 
%
\bibitem{Krein2}R.L.S. Farias, V.S. Timoteo, S.S. Avancini, M.B. Pinto, G. Krein,
Eur. Phys. J. {\bf A 53}, 101 (2017).
%
\bibitem{basar}G.~ Basar, D.~ E.~ Kharzeev, and V.~ Skokov,
Phys.\ Rev.\ Lett. {\bf 109}, 202303 (2012),
%
%
\bibitem{Li_shear}S. Li, H-U Yee,
arXiv:1707.00795 [hep-ph].
%
\bibitem{Nam_shear}S. Nam, C-W Kao,
Phys.\ Rev.\ {\bf D 87}, 114003 (2013).
%
\bibitem{Sedarkian_shear}M.~G.~Alford, H.~Nishimura and A.~Sedrakian,
Phys.\ Rev.\ C {\bf 90}, no. 5, 055205 (2014)
%
\bibitem{Tawfik_shear}A. N. Tawfik, A. M. Diab, M. T. Hussein,
Int. J. Adv. Res. Phys. Sci. 3 (2016) 4.
%
\bibitem{Tuchin_shear}K. Tuchin,
J. Phys. G: Nucl. Part. Phys. 39 (2012) 025010.
%
\bibitem{Hattori_bulk} 
K.~Hattori, X.~G.~Huang, D.~H.~Rischke and D.~Satow,
arXiv:1708.00515 [hep-ph].
%
\bibitem{Sedarkian_bulk}X-G Huang, M. Huang, D. H. Rischke, A. Sedrakian,
Phys. Rev. {\bf D 81}, 045015 (2010).
%
\bibitem{Huang_bulk} 
X.~G.~Huang, A.~Sedrakian and D.~H.~Rischke,
Annals Phys.\  {\bf 326}, 3075 (2011)
%
\bibitem{Ashutosh}P. Mohanty, A. Dash, V. Roy,
arXiv:1804.01788 [nucl-th].
%
\bibitem{Agasian_bulk1}N.O. Agasian,
Phys. Atom. Nucl. 76 (2013) 1382.
%
\bibitem{Agasian_bulk2}N.O. Agasian,
JETP Lett. 95 (2012) 171.
%
%
\bibitem{Nam_cond} 
S.~i.~Nam,
Phys.\ Rev.\ {\bf D 86}, 033014 (2012).
%
\bibitem{Hattori_cond1}K. Hattori, D. Satow,
Phys.\ Rev.\ {\bf D 94}, 114032 (2016).
%
\bibitem{Hattori_cond2}K. Hattori, S. Li, D. Satow, H.-U. Yee,
Phys.\ Rev.\ {\bf D 95}, 076008 (2017).
%
\bibitem{Sedarkian_cond} 
A.~Harutyunyan and A.~Sedrakian,
Phys.\ Rev.\ C {\bf 94}, no. 2, 025805 (2016).
%
\bibitem{Kerbikov_cond} 
B.~O.~Kerbikov and M.~A.~Andreichikov,
Phys.\ Rev.\ D {\bf 91}, no. 7, 074010 (2015)
%
\bibitem{Feng_cond} B. Feng,
Phys.\ Rev.\ {\bf D 96}, 036009 (2017).
%
\bibitem{Buividovich_cond} 
P.~V.~Buividovich, M.~N.~Chernodub, D.~E.~Kharzeev, T.~Kalaydzhyan, E.~V.~Luschevskaya and M.~I.~Polikarpov,
Phys.\ Rev.\ Lett.\  {\bf 105}, 132001 (2010).
%
\bibitem{Fukushima_cond}K. Fukushima, Y. Hidaka,
arXiv:1711.1472 [hep-ph].
%
%
%
%
%
%
%
%
%
\bibitem{Roy:2015kma}V.~Roy, S.~Pu, L.~Rezzolla and D.~Rischke,
Phys.\ Lett.\ B {\bf 750}, 45 (2015)
%
\bibitem{Pu:2016ayh}S.~Pu, V.~Roy, L.~Rezzolla and D.~H.~Rischke,
Phys.\ Rev.\ D {\bf 93}, no. 7, 074022 (2016)
%
\bibitem{Das:2016cwd}S.~K.~Das, S.~Plumari, S.~Chatterjee, J.~Alam, F.~Scardina and V.~Greco,
Phys.\ Lett.\ B {\bf 768}, 260 (2017)
%
\bibitem{Finazzo_shear}R. Critelli, S.I. Finazzo, M. Zaniboni, J. Noronha, 
Phys.\ Rev.\ D {\bf 90}, 066006 (2014);
S.~I.~Finazzo, R.~Critelli, R.~Rougemont and J.~Noronha,
Phys.\ Rev.\ D {\bf 94}, 054020 (2016).
%
\bibitem{Jain}S. Jain, R. Samanta, S. P. Trivedi,
J. High Energy Phys. 10 (2015) 028.
%
%
\bibitem{Frank}M.~Frank, M.~Buballa and M.~Oertel,
Phys.\ Lett.\ B {\bf 562}, 221 (2003).
%
\bibitem{Gavin}S. Gavin, 
Nucl. Phys. {\bf A, 435}, 826 (1985).
%
\bibitem{chakrabortty}P. Chakraborty and J. I. Kapusta, 
Phys. Rev. C {\bf 83}, 014906 (2011).
%
\bibitem{Landau}E.M. Lifshitz and L.P. Pitaevskii, 
1987 {\it Physical kinetics}, Pergamon Press, U.K.
%
%
 \bibitem{hm1}A. Abhishek, H. Mishra , S. Ghosh 
 Phys.Rev. D97 (2018)  014005.
%
 \bibitem{hm2}P. Singha, A. Abhishek, G. Kadam, S. Ghosh, H. Mishra, 
 J. Phys. {\bf G 46} (2019) 015201.
 \bibitem{hm3}P. Deb, G. P. Kadam, H. Mishra
 Phys.Rev. D94 (2016)  094002.
 \bibitem{SG_NJL1}S. Ghosh, T. C. Peixoto, V. Roy, F. E. Serna, G. Krein,
 Phys. Rev. {\bf C 93} (2016) 045205.
 \bibitem{SG_NJL2}S. Ghosh, F. E. Serna, A. Abhishek, G. Krein, H. Mishra,
 Phys.Rev. D99 (2019) 014004.
%
%
%
%
%
\end{thebibliography}
\end{document}